\documentclass[preprint,floats,aps,epsfig,nofootinbib,amssymb]{revtex4}
\usepackage{mathrsfs}

\def \pslash {\partial\!\!\!/}

\usepackage{slashed}
\usepackage{graphicx,color}
\usepackage{epsfig}
\usepackage{subfigure}
\usepackage{epsfig}
\usepackage{dcolumn}
\usepackage{bm}

\begin{document}

\title{The D0 Dimuon Charge Asymmetry and Baryon Asymmetry of the Universe}

\author{Wei Chao}
\email{chaow@pku.edu.cn}

\author{ Yong-chao Zhang}

\affiliation{$^1$Center for High Energy Physics, Peking University,
Beijing 1000871, China  }

\begin{abstract}
The D0 collaboration has reported a $3.2\sigma$ deviation from the
standard model (SM) prediction in the like-sign dimuon charge
asymmetry. New physics beyond the SM in $B_s^{} - \bar B_s^{} $
mixing is needed to explain the data.  In this paper, we investigate
the possible extension of the SM with one generation color-triplet
charged scalar as well as three generation Majorana fermions. We
study the implications of the model on the D0's dimuon charge
asymmetry as well as matter anti-matter asymmetry of the Universe.
\end{abstract}

\draft

\maketitle

\section{Introduction}
The cosmological matter-antimatter asymmetry is one of the most striking mysteries in the Universe.
The five-year observations of the WMAP colloboration precisely measured the ratio of baryon number to
photon number densities as \cite{wmap}
\begin{eqnarray}
\eta_B^{} = {n_B^{} \over n_\gamma^{} } = (6.225 \pm 0.170) \times
10^{-10}\; .
\end{eqnarray}
According to Sakharov's suggestion \cite{sarkharov}, to dynamically
generate the matter-antimatter asymmetry of the Universe, three
conditions must be satisfied, which are the violation of baryon
number conservation, the violation of C and CP, and the deviation
from thermal equilibrium. In the standard model (SM), C parity is
maximally violated and the phase of the Cabibbo-Kobayashi-Maskawa
(CKM) matrix is the sole source of the CP violation. However, this
CP violation effect, which has been precisely measured in hadronic
physics, is too small to be used to accommodate the baryon asymmetry
of the Universe. Such that, investigating the new origin of CP
violation is a subject with great significance.

More recently, the D0 collaboration \cite{d0evid}, with $6.1 {\rm
fb^{-1}}$ of data, has reported a measurement of the like-sign
dimuon charge asymmetry in semi-leptonic decay of $b$ hadrons,
\begin{eqnarray}
A_{sl}^b \equiv { N_b^{++} - N_b^{--}  \over N_b^{++} + N_b^{--} } =
- (9.57\pm 2.51 \pm 1.46) \times 10^{-3} \; ,
\end{eqnarray}
where $N_b^{++} (N_b^{--})$ is the number of events with $b (\bar
b)$ hadrons decaying semileptonically into $\mu^+ X (\mu^- X)$. This
result, with the first error being statistical and the second
systematic, is $3.2\sigma$ deviation from the SM prediction of $-0.2
\times 10^{-3}$ \cite{bbssm}. $A_{sl}^b$ was also measured by the
CDF collaboration \cite{cdf}, which has
$A_{sl}^b=(8.0\pm9.0\pm6.8)\times 10^{-3}$, using $1.6 fb^{-1}$ of
data. This result is positive but still compatible with the D0
measurement at $1.5\sigma$ level because its uncertainties are 4
times larger than those of D0. Combining in quadrature these two
results, one has $A_{sl}^b \approx - (8.5 \pm 2.8) \times 10^{-3}$
\cite{doexp8} which is still $3\sigma$ away from the SM value. If
confirmed, it will be an evidence of new physics beyond the SM.
There have already been some models \cite{doexp1, doexp8, dimuexp2}
attempting to explain the data.

The D0 collaboration's result is the first laboratory evidence for
significant matter-antimatter asymmetry. We suspect there should be
some connection between the baryon asymmetry of the Universe and
this anomaly. In this paper, we consider the possible extension of
the SM with one color-triplet charged scalar and three Majorana
fermion singlets. In this model, there are tree level processes
leading to dimension six effective operators, like ${\cal O}
(1/M_\varphi^2)\bar b s \bar c c $, which may contribute to
$B_s^{}-\bar B_s^{}$ mixing and thus the like-sign dimuon charge
asymmetry can be explained. Besides, the out of equilibrium decay of
the neutral fermion may produce $B-L$ asymmetry. This asymmetry may
be converted to the baryon asymmetry of the universe via sphaleron
process \cite{sphaleron}, which violates $B+L$ but keeps $B-L$
conservation.

The paper is organized as follows: In section II, we present our
model. Section III is devoted to investigating the like-sign dimuon
charge asymmetry. We study baryogenesis in section IV. Conclusions
and remarks are presented in section V.

\section{The model}

To reproduce the large anomalous like-sign dimuon charge asymmetry
observed by D0 and to explain the matter-antimatter asymmetry of the
Universe, new physics beyond the SM is needed. In this section, we
consider the possible extension of the SM with one charged scalar
$\varphi$ and three fermion singlets $\chi$, whose representation in
$SU(3)_C^{} \times SU(2)_L^{} \times U(1)_Y^{}$ gauge group are
$\varphi : $ $(3, 1, -1/3)$ and $\chi :$ $(1, 1, 0)$, respectively.
For simplification, $Z_3^{}$ discrete flavor symmetry is introduced
and the presentations of fields on $Z_3^{}$ are listed in table I.
The newly introduced lagrangian can be written as
\begin{eqnarray}
{\cal L}_{\rm New}^{}&=& |(\partial_\mu^{} + i \lambda_a^{} G^a_\mu
+ i Y B_\mu^{} ) \varphi |^2 + \bar \chi i \pslash \chi - {1 \over 2
} M \overline \chi \chi - {1 \over 2} M_S^{} \varphi^\dagger \varphi
- {1\over 4} \lambda_1^{}
( \varphi^\dagger \varphi ) (H^\dagger H ) \nonumber \\
&& - \zeta \overline \chi \varphi D_R^{}-\lambda U_R^T \bar \varphi
D_R^{} + {\rm h.c. } \; ,
\end{eqnarray}
where $\zeta$ and $\lambda$ are Yukawa coupling matrices, $H$ is the
SM Higgs boson, $U_R^{}$ and $D_R^{}$ represent the second or third
generation right-handed up- and down-type quarks. Assuming $\varphi$
has baryon number $2/3$, the interaction, $\bar \chi \varphi
d_R^{}$, violates baryon number by $1$ unit.
\begin{table}[htbp]
\centering
\begin{tabular}{c|c}
\hline fields & ~~~~~~$Z_3^{}$~~~~~ \\
\hline $ Q_L^{}, u_R^{}, d_R^{} $ & 1   \\
\hline $\varphi, \chi, s_R^{}, c_R^{}, t_R^{}, b_R^{} $ & $\omega$ \\
\hline$H,$ & $\omega^2$ \\
\hline
\end{tabular}
\caption{Representations of fields on $Z_3^{}$ flavor symmetry.}
\end{table}

Note that $Z_3^{}$ symmetry is explicitly broken down by the Yukawa
interactions containing the firt generation right-handed quarks and
the mass term of $\chi$.  We may introduce another  Higgs doublet,
whose  representation under $Z_3^{}$ is 1, to genreate masses for
the first generation up- or down-type quarks, and itroduce one Higgs
singlet, whose representation unde $Z_3^{}$ is $\omega$, to generate
masses for $\chi$. Such that $Z_3^{}$ symmetry can be recovered. Due
to the existence of $Z_3^{}$ symmetry, $\varphi$ and $\chi$ do not
couple to the first generation of quarks. As a result, we need not
to warry about constraints from proton stability, $K_0^{}-\bar
K_0^{}$ mixing or $D_0^{} - \bar D_0^{}$ mixing. Notice that
$\varphi$ is similar to the super-partner of $d_R^{}$ and $\lambda
u^T_R \varphi d_R^{}$ is similar to $R$-parity violating
interactions in supersymmetry. The experimental lower bound for the
mass of $\varphi$ should be consistent with that of squarks.
Integrating out heavy degrees of freedom, we can derive an effective
operator that contributes to $b\rightarrow s \gamma $. The inclusive
decay width for this process can be written as
\begin{eqnarray}
\Gamma  (b\rightarrow s \gamma )= {e^2 m_b^5 \over 16 \pi} \left( |{\cal F} |^2 + |{\cal Q}|^2 \right) \; ,
\end{eqnarray}
where ${\cal Q }$ is the contribution of SM penguin diagram and
\begin{eqnarray}
{\cal F} = {1 \over 24\pi^2 M_\varphi^2} {\lambda^*_{is}} \lambda_{ib}
 \left[ {2 \over 3(1-x) } + {1+x-4x^2 \over 2(1-x)^3 } + {2x- 3x^2
 \over (1-x)^4} \ln x \right] \; ,
\end{eqnarray}
comes from the new interaction in our model, with $x^{} = {m_i^{2} / M_\varphi^2}  $.
Assuming $\lambda_{is}^*\lambda_{ib}^{} \sim 0.1$ and $M_\varphi^{} = 500 {\rm GeV}$,
we get the predication for the branching ratio of $b\rightarrow s \gamma$: ${\rm BR} (b\rightarrow s \gamma ) = 3.53 \times 10^{-4}$,
which is consistent with its present experimental value, $(3.55 \pm 0.24^{+0.09}_{-0.10})\times 10^{-4}$,
predicted by the Heavy Flavor Average Group (HFAG) \cite{hfag}.

\section{$B_s^{} -\bar B_s^{}$ mixing}

Notice that, $A_{sl}^b$, appearing in Eq. (2), is blind as to which
flavors of $B$ meson produced the two muon, it places a constraint
on the semi-leptonic CP asymmetries of both $B_d^{}$ and $B_s^{}$
mesons, which we will call $a_{sl}^{d}$ and $a_{sl}^s$,
respectively. The relation between $A_{sl}^b$ and $a_{sl}^{s, d}$ is
given by
\begin{eqnarray}
A_{sl}^b= (0.506\pm0.043) a_{sl}^d + (0.494 \pm0.043) a_{sl}^s \; .
\end{eqnarray}
Taking into account the current experimental value of
$a_{sl}^d=-0.0047 \pm 0.0046$ \cite{abcd}, one obtains $a_{sl}^s = -0.0146 \pm
0.0075$, which is in agreement with D0's direct measurement of
$a_{sl}^{s}=-0.0017\pm0.0094$ \cite{d0evid}, albeit with large uncertainties. We
may combine all these results to obtain an average value $a_{sl}^{s}
\approx -(12.7 \pm 5.0) \times 10^{-3}$.

Theoretically, there are two amplitudes characterizing mixing in
$B_s^{}$ system: the off-diagonal element of the mass matrix
$M_{12}^s$ and the off-diagonal element of the decay matrix
$\Gamma_{12}^s$. The violation of CP is caused by the non-zero value
of the phase: $\phi_s^{} = \arg (- M_{12}^s/\Gamma_{12}^s)$. In
terms of these parameters, to a very good approximation,
$a_{sl}^{s}$ is given by \cite{cpviolation}
\begin{eqnarray}
a_{sl}^s ={ |\Gamma_{12}^s | \over  | M_{12}^s |  } \sin \phi_s^{} =
{\Delta \Gamma_s^{} \over \Delta M_s^{} } \tan \phi_s^{} \; ,
\end{eqnarray}
where $\Delta M_s^{} = 2 M_{12}^s$ and $ \Delta \Gamma_s^{} = 2
|\Gamma_{12}^s| \cos \phi_s^{}$. Notice that the SM prediction for
$\Delta M_s^{}$ agrees with data very well, while both $\Delta
\Gamma_s^{}$ and $\phi_s^{}$ differ from the experimental data by
about $1.5 \sigma $ and $3 \sigma$, respectively. It is natural to
attribute D0's result to big corrections to $\Gamma_{12}^{}$ and
$\phi_s^{}$ from new physics.

In our model, there are tree level contributions to the dimension
six effective operator ${\cal O} (1/M_\varphi^2)\bar b s \bar c c $,
which contributes to $\Gamma_{12}$ for $B_s^{} -\bar B_s^{}$ mixing.
Integrating out $\varphi$ at the tree level, we have
\begin{eqnarray}
\delta {\cal L} = { \lambda_{ik} \lambda_{jl}^* \over 2 M_\varphi^2
} ( \bar u_i^{\alpha}\gamma^\mu P_R^{} u_j^\alpha \bar d_k^\beta
\gamma_\mu^{} P_R^{} d_l^\beta - \bar u_i^{\alpha}\gamma^\mu P_R^{}
u_j^\beta \bar d_k^\beta \gamma_\mu^{} P_R^{} d_l^\alpha) \; ,
\end{eqnarray}
where $i, j, k, l$ are generation indices and $\alpha, \beta$ are
color indices. Starting with these four quark interactions, one can
derive the expression of $\Gamma_{12}^{s}$:
\begin{eqnarray}
\Gamma_{12}^{\lambda} \approx { m_b^2 \over 128 \pi m_B^{} }
{\left(\lambda_{22}^{} \lambda^{
*}_{23} \right)^2\over M_\varphi^4}  \langle Q_s^{} \rangle \; ,
\end{eqnarray}
where $ \langle Q_s^{} \rangle = \langle B_s^{} | \bar s_\alpha
\gamma_\mu^{}  P_R^{}  b_\beta^{} \bar s_\beta^{} \gamma_\mu^{} P_R
b_\alpha | \bar B_s^{} \rangle = -(5/12) f_{B_s}^2 B_{B_s}^s
m_{B_s}^2$. $B_{B_s}^s$ is the bag parameter and we take it to be
equal to one.

There are also contributions to the off-diagonal elements,
$M_{12}^s$, of the mass matrix of the $B_s^{}- \bar B_s^{}$ system
due to the box diagram. Using vacuum insertion approximation, we have
\begin{eqnarray}
M_{12}^{ \lambda} = -{1 \over 192 \pi^2 M_\varphi^2 }
(\lambda^\dagger _{3\alpha} \lambda_{\alpha 2}^{}
\lambda^\dagger_{3\beta} \lambda_{\beta 2})f_B^2 M_B^{} B_b^{}
\exp[{i( \xi_b^{}- \xi_q^{}-\xi_{B_q}^{})}] {\cal F}^* (x_\alpha^{},
x_\beta^{})\; ,
\end{eqnarray}
with \cite{cpviolation}
\begin{eqnarray}
{\cal F }(x_\alpha^{}, x_\beta^{}) &=&{1 \over
(1-x_\alpha)(1-x_\beta) } \left({7x_\alpha x_\beta \over 4} -1
\right) + {x_\alpha^2 \ln x_\alpha \over (x_\alpha-x_\beta)
(1-x_\alpha^2)} \left(1-2x_\beta + {x_\alpha x_\beta \over 4}
\right) \nonumber \\ && + {x_\beta^2 \ln x_\beta \over (x_\alpha -
x_\beta) (1-x_\beta)} \left(1-2x_\alpha + {x_\alpha x_\beta \over 4
} \right) \; , \nonumber
\end{eqnarray}
where $x_{\alpha,\beta}^{}= m_{\alpha, \beta}^{2}/ M_\varphi^2 $
($\alpha, \beta=c, t$). In deriving Eq. (10), we have ignored the
contributions from the $\zeta\chi \varphi d_R^{}$ term by assuming
${\cal O}(\zeta) \ll {\cal O} (\lambda)$.

\begin{figure}[t]
\centering
\includegraphics[height=8.5cm, width=10.5cm, angle=0]{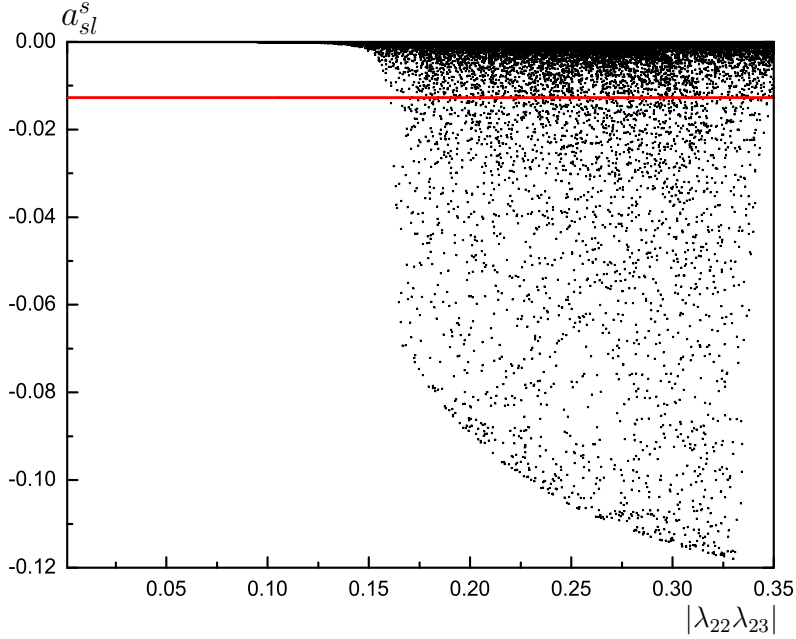}
\caption{ $a_{sl}^s $ as function of $|\lambda_{22}\lambda_{32}|$
with $M_\varphi^{}$ being random parameter in the range $[300, 600]$
(GeV). The horizontal line represents the central value of
$a_{sl}^s$ measured by $D0$.}
\end{figure}

Taking into account contributions from new physics, the final
expressions for $M_{12}^s$ and $\Gamma_{12}^s$ can be written as
\begin{eqnarray}
M_{12}^{s} = M_{12}^{\rm SM} + M_{12}^{\lambda} \; ; \hspace{1cm}
\Gamma_{12}^{s} = \Gamma_{12}^{\rm SM} + \Gamma_{12}^{\lambda} \; ,
\end{eqnarray}
where the formulae for $M_{12}^{\rm SM}$ and $\Gamma_{12}^{\rm SM}$
are given by \cite{pdgg}
\begin{eqnarray}
M_{12}^{\rm SM} &=& - {G_F^2 M_W^2 \eta_B^{} m_{B_q}^{} B_{B_q}^{}
f_{B_q}^2 \over 12 \pi^2 } (V_{tq}^* V_{tb}^{})^2  S_0^{}  \left
({m_t^2 \over M_W^2} \right ) \; ,  \\
\Gamma_{12}^{\rm SM} &=& {G_F^2 m_b^2 \eta_B^\prime m_{B_q}^{}
B_{B_q}^{} f_{B_q}^2 \over 8 \pi } \left[ (V_{tq}^* V_{tb}^{})^2 +
V_{tq}^* V_{tb}^{} V_{cq}^* V_{cb}^{} {\cal O} \left( {m_c^2 \over
m_b^2 } \right)  \right] \; .
\end{eqnarray}

Since the couplings are in general complex, we can derive arbitrary
phase $\phi_s^{}$ in general. Before giving some numerical analysis,
let's assume that the Yukawa coupling constant $\lambda$ is real,
and the operator ${\cal O} (1/ M_\varphi^2)\bar b s \bar c c$
dominates contributions to $\Gamma_{12}^\lambda$ and
$M_{12}^\lambda$. In this case, $\phi_s^{}$ as well as $a_{sl}^s$
will be functions of $|\lambda_{22} \lambda_{23}|$ and
$M_\varphi^{}$. We plot in Fig. 1, $a_{sl}^s$ as function of
$|\lambda_{22} \lambda_{23}|$ with $M_\varphi^{}$ being a random
parameter in the range $[300, 600]~ ({\rm GeV})$. The horizontal
line represents the central value of $a_{sl}^s$ measured by $D0$. It
is clear that, to fit the data, $|\lambda_{22}\lambda_{23}|$ should
be larger than $0.17$. Of course, $|\lambda_{22}^{}\lambda_{23}^*|$
can be much smaller in the case $\lambda$ being complex.

\section{Baryogenesis}

\begin{figure}[t]
\subfigure[]{\epsfig{file=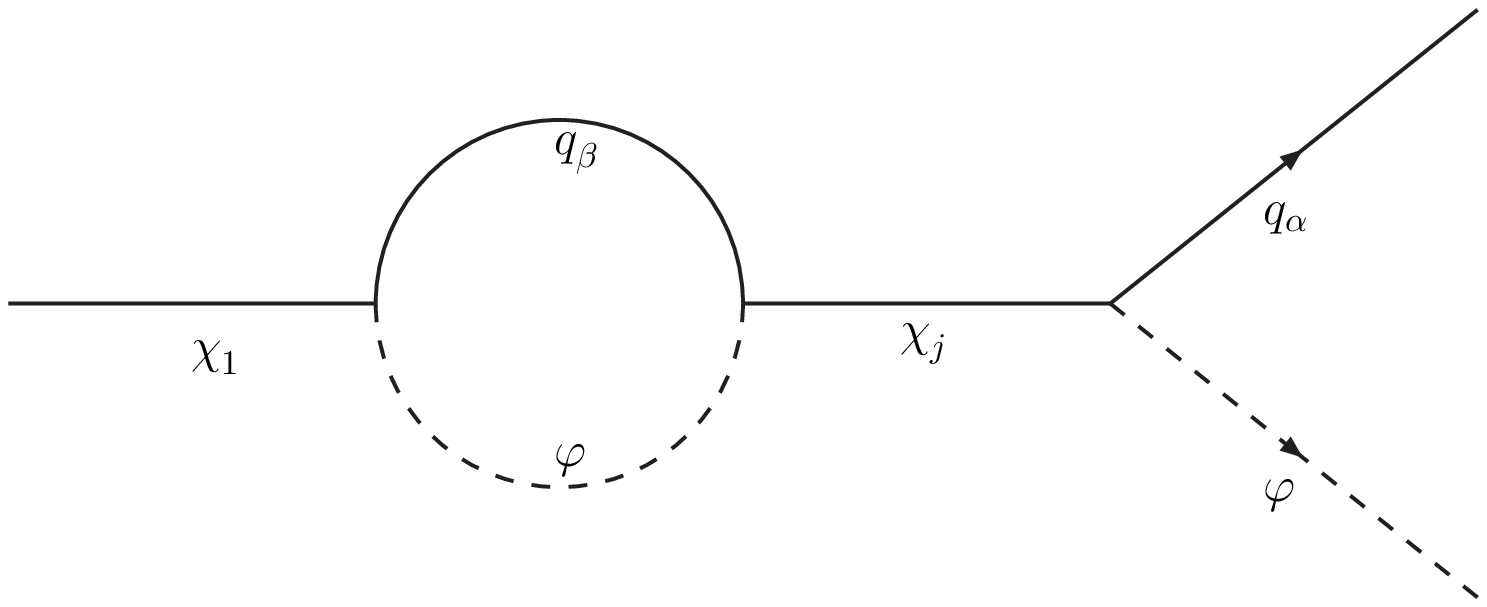,height=2.7cm,width=6.5cm,angle=0}}
\subfigure[]{\epsfig{file=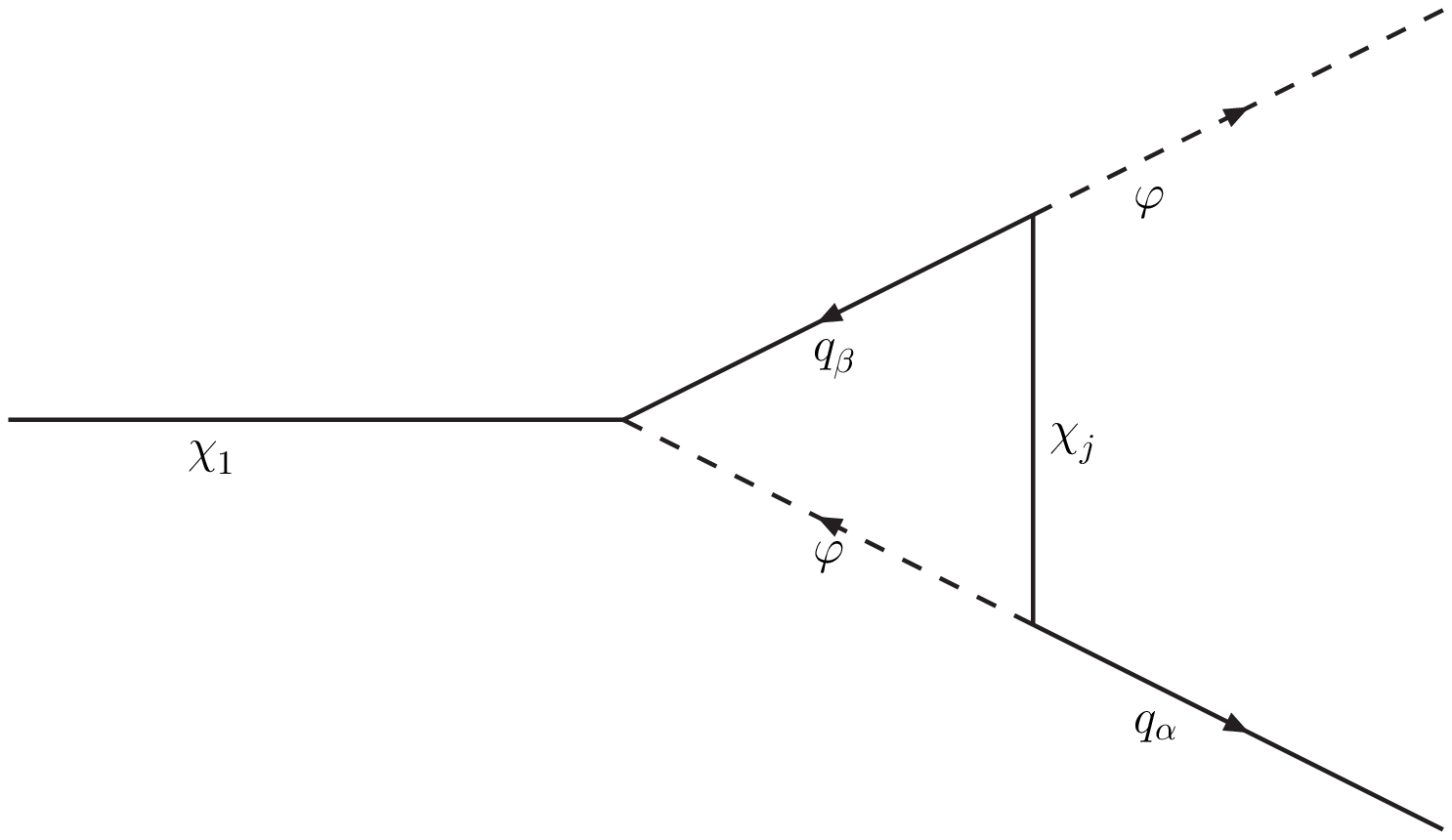,height=2.8cm,width=6.3cm,angle=0}}
\caption{The decays of $\chi$ at one loop level for generating $B-L$
asymmetry.}
\end{figure}

As noticed in the last section, the D0's like-sign dimuon charge
asymmetry can be explained by introducing a color-triplet scalar
$\varphi$. As a byproduct, this may induce baryon number violating
operators like $\zeta \bar \chi \varphi d_R^{}$, which is just what
we need to reproduce the baryon asymmetry of the universe. We assume
$\chi$ is thermally produced in the early universe. Its out of
equilibrium decay, as shown in Fig 2, may violate the $B-L$
symmetry. The expression for the CP asymmetry $\varepsilon$ in the
decay of $\chi$, which is similar to that in the decay of heavy
Majorana neutrinos in the Type-I seesaw mechanism
\cite{leptogenesis}, can be written as
\begin{eqnarray}
\varepsilon={\Gamma(\chi\rightarrow \varphi d_R^{} ) -\Gamma(\chi
\rightarrow \bar \varphi \bar d_R^{}) \over \Gamma(\chi\rightarrow
\varphi d_R^{} ) + \Gamma(\chi \rightarrow \bar \varphi \bar d_R^{})
} = { 1 \over 8 \pi } \sum_i^{} {{\rm Im } [ ( \zeta \zeta^\dagger
)_{i1}^{2} ] \over [\zeta \zeta^\dagger]_{11}^{}  } f{\left(
M_{\chi_i}^2 \over M_{\chi_1}^2 \right )} \; ,
\end{eqnarray}
where $f(x)$ is given by
\begin{eqnarray}
f(x)= \sqrt{x} \left[ { x-2 \over x-1 } -( 1 + x ) \ln \left( 1 + x
\over x \right ) \right] \; .
\end{eqnarray}

After the evolution of Boltzmann equations governing the baryon
number density, we derive the $B-L$ asymmetry stored in right-handed
quarks. Because all the quarks are in thermal equilibrium, this
asymmetry may be converted to left-handed quarks through the
left-right equilibrium \cite{diraclep}, which can be understood as
follows. Let us define the chemical potential associated with the
$q_R^{}$ field as $\mu_{qR} = \mu_0 + \mu_{BL}^{}$, where
$\mu_{BL}^{}$ is the chemical potential contributing to $B-L$
asymmetry and $\mu_0^{}$ is independent of $B-L$ . Hence at
equilibrium  we have the chemical potential associated with $Q_L^{}$
given by $\mu_{QL}=\mu_0+\mu_{BL}+\mu_H^{}$. Thus we see that the
same chemical  potential is associated with $Q_L^{}$ as that of
$q_R^{}$. Finally, the $B-L$ asymmetry is converted to the baryon
asymmetry of the Universe through the sphaleron process, which
violates $B+L$ symmetry but keeps $B-L$ conservation. Then the final
baryon asymmetry can be given as \cite{earlyun}
\begin{eqnarray}
\eta_B^{}\approx {28 \over 79} {0.3 \varepsilon \over g_*^{} K (\ln
K)^{0.6}} \; ,
\end{eqnarray}
with
\begin{eqnarray}
K= {\Gamma_D^{}\over  2 H } (T=M_\chi^{} )\approx { 3
(\zeta\zeta^\dagger)_{11}^{} M_{pl}^{} \over 32\pi \sqrt{g^{*}}
M_\chi^{}} \; ,
\end{eqnarray}
which measures the effectiveness of decays at the crucial epoch
($T\sim m_\chi^{}$) when $\chi$ must decrease in number if they are
to stay in equilibrium. Here, $M_{pl}^{} = 1.2 \times 10^{19}$ ${\rm
GeV}$ is the Planck mass and $g_*$ is effective degrees of freedom,
at the temperature where $\chi$ decouples. In deriving Eq. (16), we
have assumed that the factor $K$ is greater than 1, but not too
large. If $K\ll 1$, $\eta_B^{} \approx 28 \varepsilon/79 g_*^{}$.
For instance, inputting $M_{\chi1} = 0.1 M_{\chi 2} =10^{7} {\rm
GeV}$, K=50, $\sum_{i} (\zeta \zeta^\dagger)_{i1}^{}\sim 1.3 \times
10^{-5}$ and maximal CP-phase, we derive the sample predication:
$\varepsilon = 7 \times 10^{-5}$. In consequence, we deduce
$\eta_B^{}\sim 10^{-10}$, as desired.

\section{Concluding remarks}

We have extended the SM by introducing one color-triplet charged
scalar, $\varphi$, and three Majorana fermions, $\chi$. We have
shown that the dimuon charge asymmetry, reported by the D0
collaboration, can be explained by the Yukawa interaction, like
$\lambda U_R^T \varphi D_R^{}$. Besides, the matter-antimatter
asymmetry of the Universe can be realized by the out of equilibrium
decays of $\chi$. Notice that,  the mass of $\varphi$ can be several
hundred ${\rm GeV}$. It can be produced and detected at the Large
Hadron Collider.

\begin{acknowledgments}
The author thanks to professor Z. Z. Xing for helpful discussion.
This work was supported in part by the National Natural Science
Foundation of China.
\end{acknowledgments}

\end{document}